\documentclass[pre,twocolumn,superscriptaddress,showpacs,showkeys,amsmath,amssymb,amsfonts]{revtex4-1}

\usepackage{ulem, fullpage, amsmath, amssymb, color, verbatim, graphicx, subfigure}

\newcommand{\cmt}[1]{}

 \def\dt{\partial_t} \def\k{{\bf k}} \def\kt{{{\bf k}}} 
\def\lA{\lambda_1} \def\lB{\lambda_2} \def\d{\theta} \def\g{\gamma}
\def\mm{\sigma}

\def\Ms{\rho_{S;{\bf k}}} \def\Msr{\rho_{S;{\bf k-r_i}}} 
\def\Mn{\rho_{N;{\bf k}}} \def\Mnr{\rho_{N;{\bf k}-r_i}}
\def\Mr{\rho_{R;{\bf k}}} 

\def\Mst{{\rho}_{S;\kt}} \def\Mit{{\rho}_{I;\kt}}
\def\Mstr{{\rho}_{S;\kt-r_i}} \def\Mitr{{\rho}_{I;\kt-r_i}}

\def\ws{\Omega_{S;\k}} \def\wsr{\Omega_{S;\k-r_i}}
\def\wn{\Omega_{N;\k}} \def\wnr{\Omega_{N;\k-r_i}}
\def\wr{\Omega_{R;\k}} 
\def\wit{{\Omega}_{I;\kt}} \def\wst{{\Omega}_{S;\kt}}
\def\witr{{\Omega}_{I;\kt-r_i}} \def\wstr{{\Omega}_{S;\kt-r_i}}

\def\s{N_S} \def\n{N_N} \def\r{N_R} \def\ss{N_{\text{SS}}} \def\nn{N_{\text{NN}}} 
\def\rr{N_{\text{RR}}} \def\sn{N_{\text{SN}}} \def\rn{N_{\text{RN}}} 
\def\rs{N_{\text{RS}}} \def\rsr{N_{\text{RSR}}} \def\ssr{N_{\text{SSR}}} 
\def\nsr{N_{\text{NSR}}} \def\rnr{N_{\text{RNR}}}
\def\nnr{N_{\text{NNR}}} \def\snr{N_{\text{SNR}}}

\def\i{N_I} \def\is{N_\text{IS}} \def\ii{N_\text{II}} \def\isi{N_\text{ISI}} 
\def\ssi{N_\text{SSI}} \def\iii{N_\text{III}}  
\def\issi{N_\text{ISSI}}

\def\rhor[#1]{\rho_{S;\kt-r_{#1}}}

\begin{document}

\title{Asymptotically inspired moment-closure approximation for adaptive networks}
\author{Maxim S. Shkarayev}
\affiliation{Department of Physics and Astronomy, Iowa State University, Ames, Iowa  50011}
\author{Leah B. Shaw}
\affiliation{Applied Sciences Department, College of William \& Mary}
\begin{abstract}
Adaptive social networks, in which nodes and network structure co-evolve, are often
described using a mean-field system of equations for the density of node and link types.
These equations constitute an open system due to dependence on higher order topological
structures. We propose a new approach to moment closure based
on the analytical description of the system in an asymptotic regime. We apply the proposed
approach to two examples of adaptive networks: recruitment to a cause model and adaptive
epidemic model. We show a good agreement between the improved mean-field prediction and
simulations of the full network system.
 \end{abstract}

\date{\today}

\maketitle
\section{Introduction}

In recent years we have seen much progress in the field 
of network dynamics and dynamics on 
networks~\cite{Albert2002,Newman2003,Boccaletti2006,Arenas2008}.
Strong
interest in understanding phenomena such as disease spread on social networks, interaction online social media such as Facebook and Twitter,
dynamics of neuronal networks, and many others have encouraged development of
mathematical tools necessary to analyze the behavior of such 
systems~\cite{Goncalves2011,Josic2009,Sejnowski2001,moore2000,Pastor-Satorras2001}.

Often the first step in analyzing such systems is to represent them as 
networks, where an individual unit, e.g., a person, a user account, a neuron, 
is represented by a node, and possibility of interaction between any two 
units is represented by a link between them. The dynamical processes on such 
networks are often characterized by their statistical properties via a 
mean-field approach~\cite{Gross2006b,Shaw2008a,Shkarayev2012,Jolad2011}. Such 
mean-field equations consist of a hierarchy of equations, where the expected 
state of the nodes, due to interaction via the network, is coupled to the 
statistical description of links in the network. The dynamical evolution of 
the links in turn depends on the evolution of statistical description of node 
triples, which in turn depend on higher order structures, and so on. In other 
words, this mean-field description yields an infinite system of coupled 
equations, which usually must be truncated in order to be solvable. The 
truncated system is open and has to be closed by introducing additional 
information about the system.

The dynamics on network systems are often closed at the 
level of link equations, where the network 
information makes its first 
appearance~\cite{Murrell2004}. Perhaps the simplest 
closure approach is based on the assumption of 
homogeneous distribution of different node types in the 
system, and that the probability of finding a particular 
type of node in the neighborhood of a given node is 
independent of what else can be found in that node's 
neighborhood. This closure was shown to produce excellent 
results for many different 
systems~\cite{Keeling1997,Murrell2004,Rogers2011,Taylor2012a,Kiss2012A,Kiss2012B}. 
The 
heterogeneous mean-field approach, where conditioning on 
the total degree of nodes is introduced, may improve the 
accuracy of the approximation, although drastically 
increasing the number of equations in the 
description~\cite{Marceau2010,Pastor-Satorras2001}. 
Often, additional information about the system, such as 
the expected clustering coefficient, may be used to 
improve the closure~\cite{Taylor2012a}. In other cases, 
assumptions about the shape of degree distribution 
functions~\cite{Kiss2012B}, possibly guided by 
numerical simulations or physical 
observations~\cite{Keeling1997,Keeling2005}, may lead to 
an improvement in closure. 
Equation-free 
approaches may also be used when closing the mean-field 
equations~\cite{Reppas2012,Gross2008}.

All of the above closures often lead to a reasonable 
approximation of the system dynamics. However, they all 
suffer either from the lack of a priori knowledge of 
the validity of approximation 
or from having an excessive number of equations 
that must be analyzed. In this paper, we propose a new method that 
may lead to accurate closures and that also allows one to manage the expectation of 
the accuracy of the obtained closure. The proposed 
approach is based on simplification of the mean-field 
system of equations in some asymptotic regime. In the 
rest of the paper we demonstrate our approach by applying 
it to two adaptive network systems, i.e., networks where 
dynamical processes on the nodes affect the network 
structure, which in turn affects subsequent dynamics on 
the nodes \cite{Gross2008a}. In section~\ref{sec:NSR}, we derive a 
closure for a system modeling recruitment to a 
cause~\cite{Shkarayev2012}. In section~\ref{sec:SIS} we 
derive an improved closure for an adaptive epidemic 
model~\cite{Gross2006b}.

 \begin{figure}[tb!]
 \subfigure[]{\includegraphics[]{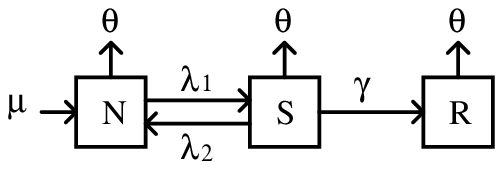}\label{fig:nsr}}
 \subfigure[]{\includegraphics[]{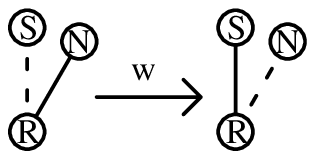}\label{fig:nsr_rew}}
 \hspace{10pt}\subfigure[]{\includegraphics[]{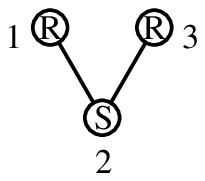}\label{fig:nsr_trp}}
 \caption{\subref{fig:nsr} Schematic representation of node dynamics in the recruitment model. The nodes are born into the N-class at rate $\mu$; nodes die at rate
$\d$; the possible transitions between the classes are marked by the arrows and
labeled with the corresponding rates ($\lA$, $\lB$, $\g$). \subref{fig:nsr_rew} Link rewiring takes place at a rate $w$.
\subref{fig:nsr_trp} Example of a node triple.
}\label{schem1}
 \end{figure}

\section{Adaptive Recruitment model} \label{sec:NSR}
 Our first example is a model for recruitment to a cause, 
introduced in~\cite{Shkarayev2012}. A society is modeled 
as a network in which some of its individuals represent a 
particular ideology and actively recruit new members. 
These nodes are referred to as the recruiting nodes, or 
R-nodes. The rest of the people in the society are either 
susceptible to recruitment or non-susceptible, referred 
to as S- and N-nodes respectively. The N-nodes may 
spontaneously change their state and become S-nodes, and 
vice versa, at rates $\lambda_1$ and $\lambda_2$, 
respectively. The R-nodes recruit S-nodes at a 
recruitment rate $\g$ per contact with S-node. A 
schematic representation of these transitions appears in 
Fig.~\ref{fig:nsr}. The R-nodes can improve their 
recruiting capability by abandoning their connections to 
N-nodes in favor of S-nodes, as shown in 
Fig~\ref{fig:nsr_rew}. This rewiring process takes place 
at a rate $w$ per contact between R- and N-nodes. The 
system is open in the sense that nodes die at rate $\d$ 
per node, and new nodes enter the system at rate $\mu$. 
The newborn nodes are born as N-nodes, and attach 
themselves with links to $\sigma$ 
randomly chosen nodes.

 In order to describe the evolution of this system, we 
begin with developing a heterogeneous mean-field 
description~\cite{Marceau2010}. We characterize 
the time evolution of $\rho_{\alpha;\k}$, the expected 
number of nodes of type $\alpha$ with $k_1$, $k_2$, and 
$k_3$ neighbors of type N, S, and R, respectively, in 
their neighborhoods, where $\k=(k_1,k_2,k_3)$:
 \begin{subequations}
 \begin{align}
 \begin{split}
 &\dt \Mn =\\
 &\lB \Ms-\lA \Mn - \d \Mn+\mu \delta_{k_1+k_2+k_3,\sigma}\\
 &+\sum_{i}\left[ \wnr(r_i) \Mnr- \wn(r_i) \Mn\right],
 \end{split}\\
 \begin{split}
 &\dt \Ms =\\
 &-\lB \Ms-\g k_3 \Ms+\lA \Mn - \d \Ms+\\
 &+\sum_{i}\left[ \wsr(r_i) \Msr- \ws(r_i) \Ms \right].
 \end{split}\\
 \begin{split}
 &\dt \Mr =\g k_3 \Ms-\d \Mr+\\
 &+\sum_{i}\left[ \wsr(r_i) \Msr- \ws(r_i) \Ms \right].
 \end{split}
 \end{align}\label{eq:master_RSR}
 \end{subequations} The allowed transitions and the 
corresponding rates shown in the Table~\ref{table:rsr}. 
In the recruiting transition rates listed in the 
table, function $P$ (function $Q$) corresponds to the 
expected number of node chains that originate 
at a given N-node (S-node) with a neighborhood 
specified by $\k$ that is connected to an S-node, which 
is turn is connected to an R-node. The terms 
$N_{\text{X}_1\ldots \text{X}_n}$ correspond to the 
expected number of node chains in the system, where a 
node chain constitutes a set of nodes, connected as 
follows: a node of type $X_1$ is connected to the node of 
type $X_2$, which in turn is connected to node of type 
$X_3$ etc. 
For example, 
$\s$ is the expected number of S-nodes in the network, 
while $\rs$ is the expected number of links with S- and 
R-nodes at its ends. In our definition of node chains we 
require the $i$th and $i+1$st nodes to be different; 
however, $i$th and $i+2$nd nodes can in fact be 
the same node. In the example of a network presented in 
Fig.~\ref{fig:nsr_trp} there are 4 RSR triples, 
corresponding to the following node combinations: 1-2-1, 
1-2-3, 3-2-1, 3-2-3. Note that the order in which nodes 
appear matters, which, for example, means that 
$\ss$ corresponds to the twice the expected number of 
undirected links between two susceptible nodes. 

\begin{table}
\caption{
Transitions and nonzero transition rates in} Eq.~(\ref{eq:master_RSR})
\centering
\begin{tabular}{l|l}
\hline\hline
transition & rate\\
\hline
$r_1=(-1,1,0)$ & $\wn(r_1)=\ws(r_1)=\wr(r_1)=\lA k_1$\\
$r_2=(1,-1,0)$ & $\wn(r_2)=\ws(r_2)=\wr(r_2)=\lB k_2$\\
$r_3=(-1,0,0)$ & $\wn(r_3)=\ws(r_3)=\wr(r_3)=\d k_1$\\
$r_4=(0,-1,0)$ & $\wn(r_4)=\ws(r_4)=\wr(r_4)=\d k_2$\\
$r_5=(0,0,-1)$ & $\wn(r_5)=\ws(r_5)=\wr(r_5)=\d k_3$\\
$r_6=(1,0,0)$ & $\wn(r_6)=\ws(r_6)=\wr(r_6)=$\\
		& $=\sigma \mu/(\n+\s+\r)$\\
$r_7=(0,-1,1)$ & $\wn(r_7)=\g P(\k)$, $\ws(r_7)=\g Q(\k)$\\
$r_8=(0,0,-1)$ & $\wn(r_8)=w k_3$\\
$r_9=(0,0,1)$ & $\ws(r_9)=w\rn/\s$\\
$r_{10}=(-1,1,0)$ & $\wr(r_{10})=w\rn/\s$\\
\hline
\end{tabular}
\label{table:rsr}
\end{table}

 The heterogeneous mean-field equations are high 
dimensional and, therefore, are extremely difficult to 
analyze. A common way to analyze the dynamics of social 
networks is via lower dimensional mean-field 
equations. These can be generated by multiplying the 
heterogeneous mean-field equations by 
$k_1^{i_1}k_2^{i_2}k_3^{i_3}$ for some non-negative 
integer values of $i_j$, and summing over $\k$. Thus, the 
equations describing node dynamics are obtained by taking 
$i_1+i_2+i_3=0$, as given in 
Eqs.~(\ref{eq:node_N})-(\ref{eq:node_R}) of Appendix~\ref{app:NSR}, while the 
description of the link dynamics is obtained by taking 
$i_1+i_2+i_3=1$, as given in 
Eqs.~(\ref{eq:link_NN})-(\ref{eq:link_RR}).  

 The hierarchy of equations generated in this manner must 
be truncated in order to obtain a finite dimensional 
description of the system. Such truncation leaves the 
system open and in need of closure. For example, the 
system of node and link equations in~(\ref{eq:NSR_MF}) 
contains the terms $\nsr$, $\ssr$ and $\rsr$, which are 
higher order structures.  The usual approach to closure 
comes from the assumption of homogeneous distribution of 
the R-nodes in the neighborhood of susceptible nodes, which leads 
to the following closure equations:
 \begin{subequations}
 \begin{align}
 &\frac{N_\text{XSR}}{\s}=\frac{N_{\text{XS}}}{\s}\frac{\rs}{\s},\label{eq:closure_XSR}\\
 &\frac{\rsr}{\s}=\left(\frac{\rs}{\s}\right)^2+\frac{\rs}{\s},\label{eq:closure_RSR_old}
 \end{align}
 \end{subequations}
 where we also assumed that the degree distribution of 
susceptible nodes is Poisson. The details of these closures are presented in the 
Appendix~\ref{app:closure}. These closures are ad hoc and 
may fail to capture the system behavior accurately if, 
for example, correlations are present. Here we develop an 
approach that derives the closure based on the system 
behavior in some asymptotic regime. In particular, we 
derive the equations that describe the evolution of node 
triples, take a steady-state relation and consider it in 
the asymptotic regime, where we are able to close the 
equations. Finally, we numerically explore the 
performance of the derived closures in 
parameter regimes outside of the asymptotic limit and 
outside of the steady-state.

 \subsection{Closing of NSR and SSR terms}
 We develop a closure of the $\nsr$ and $\ssr$ terms by considering the 
evolution of the expected number of node triples in the limit of 
$\g,\d,\mu/(\n+\s+\r) \ll w, \lA, \lB$. We consider the following expression:
 \begin{align}
 \begin{split}
 &\sum_{\k}\left(\frac{\rs}{\s}\frac{\dt\Ms}{\n}+\frac{\rn}{\n}
\frac{\dt\Mn}{\n}\right)\left(k_1 k_3 + k_2 k_3\right).
 \end{split}
 \end{align}
 This relation is evaluated at the steady state and using
Eq.~(\ref{eq:nsr_limit1}) and~(\ref{eq:nsr_limit2}).
After some algebraic manipulations described in Appendix~\ref{app:NSRSSR},
the above relation leads to
 \begin{align}
 \begin{split}
 &\frac{\nsr}{\s}+\frac{\ssr}{\s}=\frac{\sn}{\s}\frac{\rs}{\s}+\frac{\ss}{\s}\frac{\rs}{\s}
 \end{split}\label{eq:SSR_NSR_closure}
 \end{align}
 a result that is consistent with but does not imply the closure in 
Eqs.~(\ref{eq:closure_XSR}) for $X=N$ and $S$.

\begin{figure}[tb!]
 \subfigure[$\ssr+\nsr$ $\lA=10^{1}, \lB=10^{2}$]{\includegraphics[]{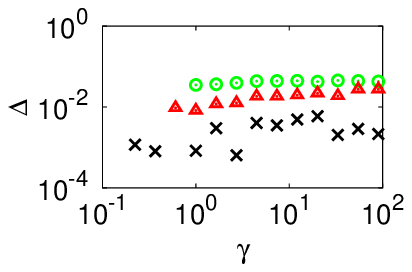}\label{fig:SSR_NSR_11}}
 \subfigure[$\ssr$  ~~$\lA=10^{1},\lB=10^{2}$]{\includegraphics[]{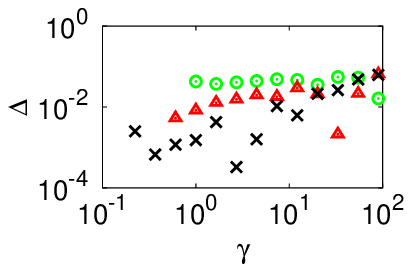}\label{fig:SSR_11}}
 \subfigure[$\ssr+\nsr$ $\lA=10^{-1}, \lB=10^{0}$]{\includegraphics[]{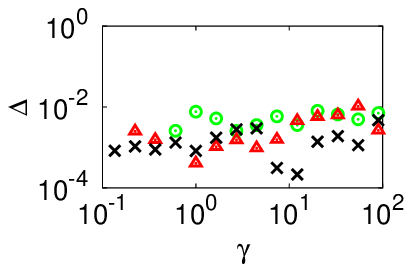}\label{fig:SSR_NSR_9}}
 \subfigure[$\ssr$ $\lA=10^{-1}, \lB=10^{0}$]{\includegraphics[]{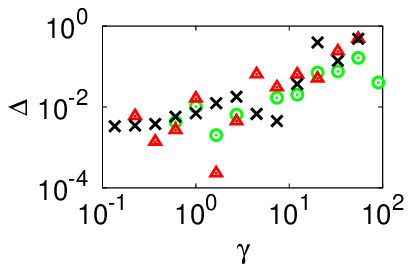}\label{fig:SSR_9}}
 \caption{Relative errors in $\nsr+\ssr$ closure 
(Eq.~(\ref{eq:SSR_NSR_closure})) (panels a,c) and $\ssr$ 
closure (Eq.~(\ref{eq:closure_XSR})) (panels b,d) 
at steady 
state, as a function of $\g$ for $w=10^{0}$ (circle, 
green online), $w=10^{1}$ (triangle, red online), 
$w=10^{2}$ (cross, black online). The simulations are 
performed following the continuous time algorithm 
introduced in~\cite{Gillespie}. The other parameters are 
$\d=1$, $\sigma=10$, $\mu=10^5$.
}\label{fig:NSR_SSR}
 \end{figure}

 We compare the asymptotically derived result of 
Eq.~(\ref{eq:SSR_NSR_closure}) and the ad hoc closure for 
the $N_{\text{SSR}}$ term in Eq.~(\ref{eq:closure_XSR}) 
with the corresponding values measured in the Monte Carlo 
simulations. Figure~\ref{fig:NSR_SSR} presents the 
relative error of the two closures
 \begin{align}
 \Delta = \left|1-\frac{\text{approximation}}{\text{exact value}} \right| \label{eq:delta}
 \end{align}
 where simulation measurements are used as the exact 
value. We can see in Fig.~\ref{fig:SSR_NSR_11} that the 
expected number of NSR and SSR triples per susceptible 
node, $\nsr/\s+\ssr/\s$, is well approximated (error on the order of 
about 1\% or less) by the 
closure of Eq.~(\ref{eq:SSR_NSR_closure}) in the large $\lA$ and 
$\lB$ limit, further improving as $w$ is increased.
According 
to Fig.~\ref{fig:SSR_NSR_9}, the closure of 
$\nsr/\s+\ssr/\s$ continues to hold even in the parameter 
regime outside of the considered limit. As for the 
individual closures, Fig.~\ref{fig:SSR_11} shows that the 
closure of $\ssr/\s$ holds as well in the large $\lA$ and 
$\lB$ regime. However, as we can see in 
Fig.~\ref{fig:SSR_9}, the closure 
fails for small $\lA$ and $\lB$, especially as $\g$ 
becomes dominant. Since the closure of $\ssr/\s+\nsr/\s$ 
is derived in the asymptotic regime, we expect it to 
be accurate at least in that limit, while $\ssr/\s$ 
closure is still ad hoc, and, therefore, deviation from 
simulations is not unexpected.

 \subsection{Closure of RSR term}
 In order to develop a closure for the $\rsr$ term, we consider the expression
 \begin{align}
 \begin{split}
 &\sum_{\k}k_3^2\dt\Ms,
 \end{split}
 \end{align}
 which leads to the equation describing the evolution of 
the expected number of RSR triples:
 \begin{align}
 \begin{split}
 &\dt \rsr = - \lB \rsr - \g \sum_{\k} k_3^3 \Ms (\k) +\\
 & \lA \rnr-\d \rsr+\g(2N_\text{RSSR}+\ssr)+\\
 &w \frac{\rn}{\s} (2\rs+\s)+\d(-2 \rsr + \rs).\label{eq:rsr_time}
 \end{split}
 \end{align}
Analyzing the steady state of this equation in the limit where
$\g,\d,\mu/(\n+\s+\r) \ll w, \lA, \lB$, and utilizing the relations in Eqs.~(\ref{eq:nsr_limit1})
and~(\ref{eq:nsr_limit2}), we are able to solve this equation and obtain the
following relation:
 \begin{align}
 \frac{\rsr}{\s} = \left(\frac{\rs}{\s}-\frac{\rn}{\n}\right)\left(2\frac{\rs}{\s}+1\right)+\frac{\rnr}{\n}.
\label{eq:rsr}
 \end{align}
 Note that, unlike the ad hoc closure of Eq.~(\ref{eq:closure_RSR_old}), this result should at least be
accurate in the considered limit, and, therefore should be more reliable.

 In order to close the $\rnr/\n$ term in 
Eq.~(\ref{eq:rsr}) we analyze the limiting behavior of 
the following expression:
 \begin{align}
 \begin{split}
 &\sum_{\k}k_3^2[\dt\Ms + \dt\Mn],
 \end{split}
 \end{align}
which in steady state reduces to:  
 \begin{align}
 &\frac{\rnr}{\n}=\frac{\rn}{\n}+\frac{\rn}{\n}\frac{\rs}{\s}\label{eq:rnr_closure}
 \end{align}
 Upon substituting the result of Eq.~(\ref{eq:rnr_closure}) into Eq.~(\ref{eq:rsr}) we obtain the following closure of the $\rsr/\s$ term:
 \begin{align}
 \frac{\rsr}{\s} =
2\left(\frac{\rs}{\s}\right)^2+\frac{\rs}{\s}-
\frac{\rs}{\s}\frac{\rn}{\n}.\label{eq:closure_RSR_new}
 \end{align}

\begin{figure}[tb!]
 \subfigure[$\lA=10^{1}, \lB=10^{2}$ $w=10^2$]{\includegraphics[width=1.54in]{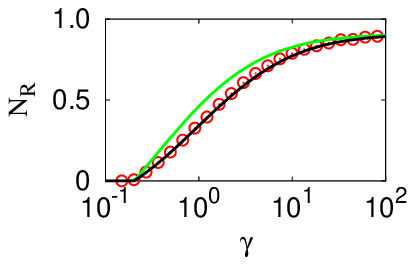}\label{fig:NSR_100_3}}
 \subfigure[$\lA=10^{1}, \lB=10^{2}$ $w=10^2$]{\includegraphics[width=1.54in]{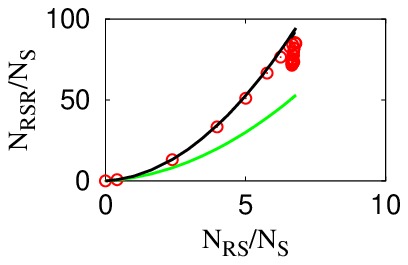}\label{fig:NSR_RSR_100_3}}
 \subfigure[$\lA=10^{1}, \lB=10^{2}$ $w=10^{-1}$]{\includegraphics[width=1.54in]{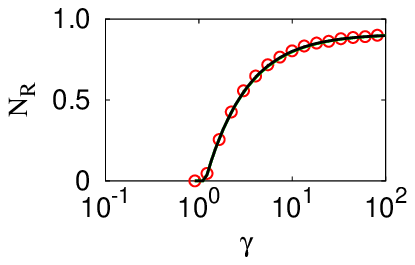}\label{fig:NSR_100_0}}
 \subfigure[$\lA=10^{1}, \lB=10^{2}$ $w=10^{-1}$]{\includegraphics[width=1.54in]{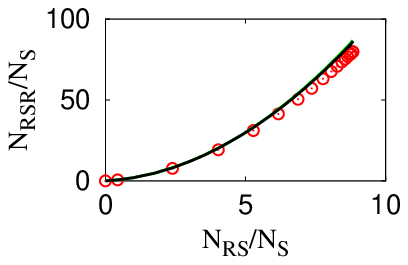}\label{fig:NSR_RSR_100_0}}
 \subfigure[$\lA=10^{-1}, \lB=10^{0}$ $w=10^{2}$]{\includegraphics[width=1.54in]{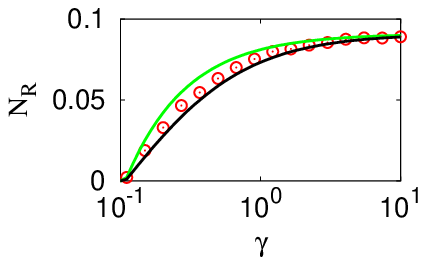}\label{fig:NSR_1_3}}
 \subfigure[$\lA=10^{-1}, \lB=10^{0}$ $w=10^{2}$]{\includegraphics[width=1.54in]{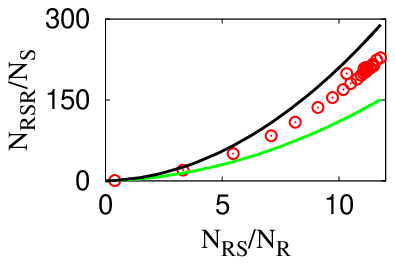}\label{fig:NSR_RSR_1_3}}
 \subfigure[$\lA=10^{-1}, \lB=10^{0}$ $w=10^{-1}$]{\includegraphics[width=1.54in]{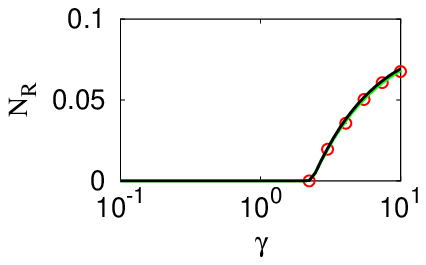}\label{fig:NSR_1_0}}
 \subfigure[$\lA=10^{-1}, \lB=10^{0}$ $w=10^{-1}$]{\includegraphics[width=1.54in]{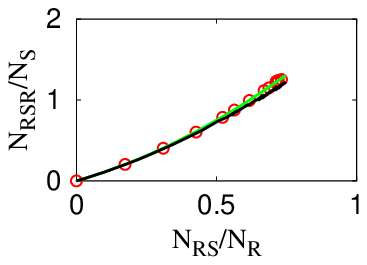}\label{fig:NSR_RSR_1_0}}
 \caption{Recruitment level, $\r$, and expected number of 
RSR triples per S-node, $\rsr/\s$, as a function of 
recruitment rate $\g$, for several sets of parameters 
$\lA$, $\lB$ and $w$. Simulation results are shown by 
circles (red online). In \subref{fig:NSR_100_3}, 
\subref{fig:NSR_100_0}, \subref{fig:NSR_1_3} 
and~\subref{fig:NSR_1_0}, the curves correspond to 
solution of mean-field equations, while in 
\subref{fig:NSR_RSR_100_3}, \subref{fig:NSR_RSR_100_0}, 
\subref{fig:NSR_RSR_1_3} and~\subref{fig:NSR_RSR_1_0} the 
curves correspond to the approximation of $\rsr/\s$ using 
two different closures. Dark curves (black online) 
correspond to closure in Eq.~(\ref{eq:closure_RSR_new}), 
while light curves (green online) correspond to closure 
in Eq.~(\ref{eq:closure_RSR_old}). The other parameters 
are same as in Fig.~\ref{fig:NSR_SSR}. Note that in 
Fig.~\subref{fig:NSR_100_0},~\subref{fig:NSR_RSR_100_0},~\subref{fig:NSR_1_0}, 
and~\subref{fig:NSR_RSR_1_0} the curves corresponding to 
the two analytic solutions lie on top of each other.
}\label{fig:NSR_RSR}
 \end{figure}

 In Fig.~\ref{fig:NSR_RSR}, we compare the performance of 
the new closure of $\rsr$ in 
Eq.~(\ref{eq:closure_RSR_new}) to the ad hoc, homogeneity 
based closure of Eq.~(\ref{eq:closure_RSR_old}). We 
consider the numerical solution of the mean-field 
equations, found in Appendix~\ref{app:NSR}, closed 
according to the two methods, and 
compare those to the steady-state size of the 
recruiting class measured in the direct network 
simulations. In both cases, the $N_\text{NSR}$ and 
$N_\text{SSR}$ terms are 
closed according to the homogeneity assumption in 
Eq.~(\ref{eq:closure_XSR}). Thus, in 
Fig.~\ref{fig:NSR_100_3} we see that in the considered 
limit, i.e., when $\lA, \lB$ and $w$ are large, the 
mean-field closed using our approach is in much better 
agreement with the simulations than the ad hoc assumption 
of Eq.~(\ref{eq:closure_RSR_old}). The reason for the 
superior performance lies in the better approximation of 
the $\rsr/\s$ term, shown in 
Fig.~\ref{fig:NSR_RSR_100_3}. Here $\rsr/\s$ and 
$\rs/\s$ are parametrized by $\g$, with larger values of 
$\rs/\s$ corresponding to the larger values of $\g$. 
Notice that the performance of the closure is reduced at 
the larger values of $\g$, as the system moves outside of 
the considered limit.

The appeal of this approach is evident when we test it outside of the 
derivation limit. In Figs.~\ref{fig:NSR_100_0} and~\ref{fig:NSR_RSR_100_0} we 
see that, when we reduce $w$, the mean-field recruited fraction and the RSR 
closure continue to be in a good agreement with the simulations. We also note 
that in this limit the new closure approaches the homogeneity closure. We 
note that when the rewiring is slow relative to transitions between N and S, 
the expected number of R neighbors should be similar for the two node types. 
This would make the last term in Eq.~(\ref{eq:closure_RSR_new}) approach 
$(\rs/\s)^2$, explaining why the two closures are close. In 
Fig.~\ref{fig:NSR_1_3}, as $\lA$ and $\lB$ are reduced, the new 
mean-field solution appears to be less consistent with the simulations, which 
is also reflected in the closure in Fig.~\ref{fig:NSR_RSR_1_3}. Finally, in 
Fig.~\ref{fig:NSR_1_0} all of the parameters are about the same order, and 
yet the asymptotically derived closure and the corresponding mean-field 
are very much consistent with the simulations.

\begin{figure}[tb!]
 \subfigure[]{\includegraphics[]{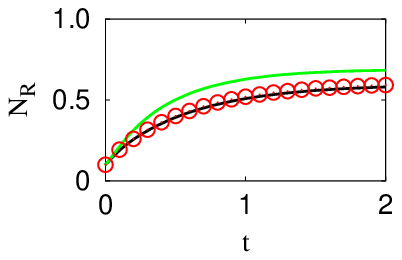}\label{fig:NSR_time}}
 \subfigure[]{\includegraphics[]{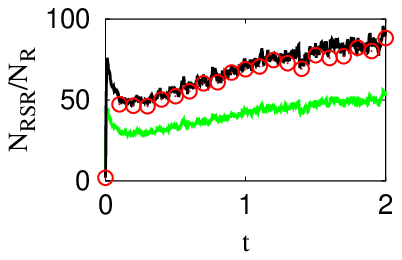}\label{fig:NSR_time_triple}}
 \caption{ Figure~\subref{fig:NSR_time} contains 
measurement and approximation of $\r$. The circles (red 
online): simulation results, the light curve (green 
online): mean-field with homogeneous closure, the dark 
curve (black online): mean-field with asymptotically 
developed closure from 
Eq.~(\ref{eq:closure_RSR_new}). 
Figure~\subref{fig:sis_time_triple} shows the time 
evolution of the number of RSR triples per S-node 
(circles, red online), the approximate value obtained 
from the relation in Eq.~(\ref{eq:closure_RSR_old}) 
(light curve, green online) and from 
Eq.~(\ref{eq:closure_RSR_new}) (dark curve, black 
online). The simulations are performed with $w=10^2$, 
$\lA=10^1$, $\lB=10^2$ and $\g=3.0$. The system evolves 
from a realization of Erd\"{o}s-R\'{e}nyi network, with mean degree $10$ and 
$10^{5}$ nodes, 85{\%} of which are N-nodes, 5\% S-nodes and 10\% 
R-nodes. The results are averaged over 10 dynamical 
realizations.
}
 \end{figure}

Thus far we have shown that our method has produced a 
closure that is a good match for the simulated system in 
steady-state, and is either superior to or as good as the 
ad hoc homogeneity closure. We further test the 
performance of our closure by using it outside of 
steady-state. Figures~\ref{fig:NSR_time} 
and~\ref{fig:NSR_time_triple} show that our closure 
continues to be consistent with the simulations even 
during the transient period. This suggests that the time 
derivative of $\rsr$ in Eq.~(\ref{eq:rsr_time}) can
be neglected in the considered limit.

 \section{Adaptive Epidemic model}\label{sec:SIS}
\begin{figure}[tb!]
 \subfigure[]{\includegraphics[]{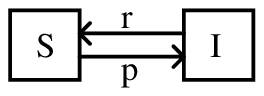}\label{fig:sis}} \hspace{20pt}
 \subfigure[]{\includegraphics[]{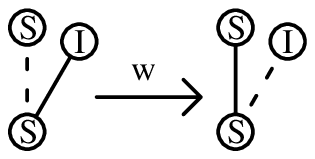}\label{fig:sis_rew}}
 \caption{Schematic representation of \subref{fig:sis} node dynamical rules and \subref{fig:sis_rew} link
dynamical rules in the adaptive epidemic model.
}\label{schem2}
 \end{figure}

 The other example that we consider is a model for epidemic spread on an adaptive social
network~\cite{Gross2006b}.  Here the disease spread is described using
the susceptible-infected-susceptible model, where each individual in the society
is in one of the two states: sick or {\it infected}, and healthy but {\it
susceptible} to infection. In the framework of networks, we refer to these as I-
and S-nodes respectively. The infected individuals become susceptible at recovery rate
$r$. The disease can spread at a rate $p$ from infected individuals to
susceptible ones via a contact between them, where the existence of the contact
is defined by the network structure. The adaptation mechanism allows
susceptible individuals to change their local connectivity to avoid
contact with infected individuals. Thus, the susceptibles rewire their contacts
away from infecteds at rate $w$, connecting instead to a randomly chosen susceptible. The node and link
dynamical rules are summarized in the Fig.~\ref{fig:sis} and~\ref{fig:sis_rew} respectively.

The evolution of the ensemble average of such a system is described by
the set of heterogeneous mean-field equations:
 \begin{subequations}
 \begin{align}
 \begin{split}
 &\dt \Mst = r\Mit-p k_2 \Mst+\\
 &+\sum_{i} \left[\wstr(r_i) \Mstr- \wst(r_i) \Mst\right],
 \end{split}\label{eq:masterS}\\
 \begin{split}
 &\dt \Mit = -r\Mit+p k_2 \Mst+\\
 &+\sum_{i}\left[ \witr(r_i) \Mitr- \wit(r_i) \Mit\right],
 \end{split}\label{eq:masterI}
 \end{align}\label{eq:master_SIS}
 \end{subequations}
 where the value of $\Mst$ (value of $\Mit$) corresponds 
to the number of S-nodes (I-nodes) with $k_1$ of S-nodes 
and $k_2$ of I-nodes in their neighborhoods, with 
$\kt\equiv (k_1,k_2)$. The function $K$ (function $M$) 
corresponds to the expected number of node chains that 
originate at an S-node (I-node) with a neighborhood 
specified by $\k$, which connects to an S-node, which in 
turn connects to an I-node. The functions 
$N_{\text{X}_1\ldots \text{X}_n}$ are defined in the same 
way as in Section~\ref{sec:NSR}.

\begin{table}
\caption{
Transitions and nonzero transition rates in
Eq.~(\ref{eq:master_SIS}).}
\centering
\begin{tabular}{l|l}
\hline\hline
transitions & non-zero rates\\
\hline
$r_1=(1,-1)$ & $\wit(r_1)=\wst(r_1)=r k_2$ \\
$r_2=(-1,1)$ & $\wst(r_2)=p K(\kt)$, $\wit(r_2)=p M(\kt)$\\
$r_3=(1,-1)$ & $\wst(r_3)=w k_2$\\
$r_4=(1,0)$ & $\wst(r_4)=w \is/\s$\\
$r_5=(-1,0)$ & $\wit(r_5)=w k_1$\\
\hline
\end{tabular}
\label{table:sis}
\end{table}

 The mean-field equations are generated by multiplying 
the heterogeneous mean-field equations 
by $k_1^{i_1}k_2^{i_2}$ and summing over $\k$, where 
$i_1$ and $i_2$ are nonnegative integers. Thus, two node 
equations are generated for $i_1+i_2=0$, and three 
distinct link equations are generated for $i_1+i_2=1$. 
These equations, presented in the appendix as 
Eqs.~(\ref{eq:node_S2})-(\ref{eq:node_II}), are open due 
to dependence on terms describing the expected 
number of ISI triples, $\isi$, and SSI triples, $\ssi$. 
In order to close this system of equations, additional 
information is required. Once again, the usual 
approach~\cite{Keeling1997} is to make an assumption 
that infected nodes are homogeneously distributed in 
the neighborhood of S-nodes, an assumption that leads to 
the following closure:
 \begin{subequations}
 \begin{align}
 &\frac{\ssi}{\s}=\frac{\ss}{\s}\frac{\is}{\s},\label{eq:closure_SSI_old}\\
 &\frac{\isi}{\s}=\left(\frac{\is}{\s}\right)^2+\frac{\is}{\s},\label{eq:closure_ISI_old}
 \end{align}
 \end{subequations}
 where we make an additional assumption that the total 
degree distribution of susceptible nodes is Poisson. 
More details can be found in 
Appendix~\ref{app:closure}.

 \begin{figure}[tb!]
 \includegraphics[]{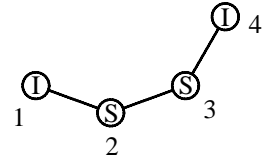}
 \caption{Schematic representation of four-node chains I-S-S-I. The term $\issi$ corresponds to the
expected number of such chains.}\label{fig:issi}
 \end{figure}

We derive a new closure of the ISI term by first considering 
the evolution of the number of ISI triples. Multiplying 
equation~(\ref{eq:masterS}) by $k_2^2$ and 
summing over $\k$ at steady state, 
 \begin{align}
 \begin{split}
 \sum_{\k} k_2^2 \dt\Mst = 0,
 \end{split}
 \end{align}
 we obtain the following equation:
 \begin{align}
 \begin{split}
 & 0=r \iii - p \sum_{\k} [k_2^3 \Mst] +\\
 & +(r+w)[-2\isi+\is] +p [2\issi+\ssi], \label{eq:isi_time}
 \end{split} 
 \end{align}
where the four-point term ISSI corresponds to the total 
number of node configurations shown in 
Fig.~\ref{fig:issi}. Using the steady-state relations in 
Eqs.~(\ref{eq:ss1}) and~(\ref{eq:ss2}), we arrive at
 \begin{align}
 \begin{split} 
&2(r+w)\s\left(\frac{\isi}{\s}-\frac{\is}{\s}-\frac{\is}{\s}\frac{\issi}{\ssi}\right)=\\ 
&=r \i\left(\frac{\iii}{\i} - \frac{\ii}{\i}\frac{\sum_{\k} [k_2^3 \Mst]}{\sum_{\k} [k_2^2 \Mst] }\right).\label{eq:isi_first}
 \end{split}
 \end{align}
 The left hand side of the equation corresponds 
to the flux of the expected number of ISI 
triples due to the changes in the neighborhood 
of the susceptible nodes, while the right hand 
side corresponds to the flux due to the 
infection and recovery of the susceptible node in the ISI 
triple. In the limit of large infection rate and 
weak rewiring, the amount of time any node 
spends in the susceptible state approaches zero. 
Therefore, it is reasonable to assume that the 
flux of triples due to the changes in the 
neighborhood of the susceptible node will 
approach zero as well. This leads us to conclude 
that the two sides of Eq.~\ref{eq:isi_first} 
must vanish, leaving us with the following 
relation:
 \begin{align}
 \frac{\isi}{\s}= \frac{\is}{\s}+\frac{\is}{\s}\frac{\issi}{\ssi}\label{eq:isi}
 \end{align}
Finally, we note that the term $\issi/\ssi$ corresponds 
to the expected number of I-nodes, node 1 in 
Fig.~\ref{fig:issi}, attached to the chain of nodes 
numbered 2, 3 and 4 in that figure. This relation is 
well approximated by the homogeneity assumption, that the 
information about the neighborhood of the 3rd node in 
Fig.~\ref{fig:issi} has no effect on the information 
about the neighborhood of the 2nd node. In other words, 
the following moment closure is considered:
 \begin{align}
 &\frac{\issi}{\ssi}=\frac{\ssi}{\ss}.
 \end{align}
 In other words, we make a homogeneity 
assumption about the neighborhood of a neighbor, 
and we expect this assumption to be more 
accurate than the same assumption about a given 
node's neighborhood, i.e., the closures in 
Eqs.~(\ref{eq:closure_SSI_old}) 
and~(\ref{eq:closure_ISI_old}).

 \begin{figure}[tb!]
 \subfigure[~$p/r=10^{-1/2}$]{\includegraphics[]{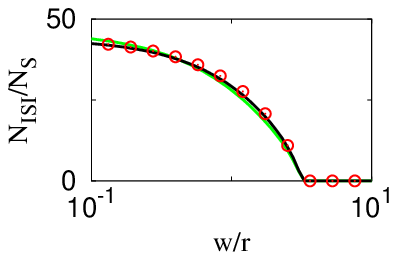}\label{fig:ISI_W1}}
 \subfigure[~$p/r=10^{-1/2}$]{\includegraphics[]{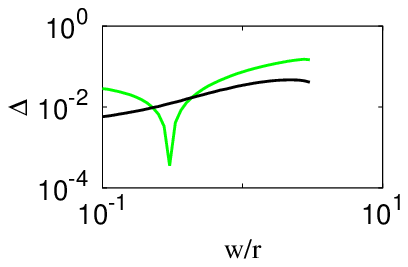}\label{fig:ISI_WD1}}
 \subfigure[~$p/r=10^{0}$]{\includegraphics[]{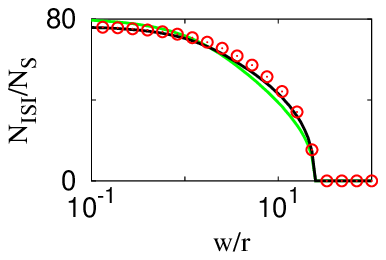}\label{fig:ISI_W2}}
 \subfigure[~$p/r=10^{0}$]{\includegraphics[]{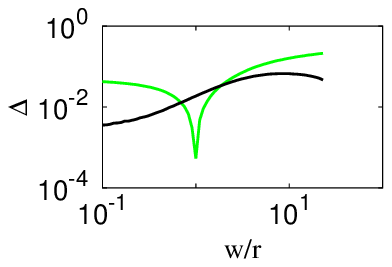}\label{fig:ISI_WD2}}
 \subfigure[~$p/r=10^{1}$]{\includegraphics[]{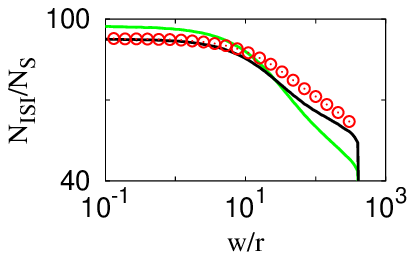}\label{fig:ISI_W3}}
 \subfigure[~$p/r=10^{1}$]{\includegraphics[]{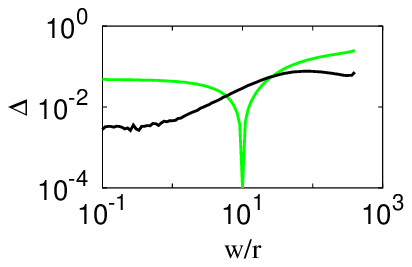}\label{fig:ISI_WD3}}
 \caption{Number of ISI triples per S-node as a function 
of rewiring rate, for several infection rates: 
simulations compared to the moment closures of 
Eq.~(\ref{eq:closure_ISI_old}) and 
Eq.~(\ref{eq:closure_ISI_new}). 
Figures~\subref{fig:ISI_W1}, \subref{fig:ISI_W2}, 
and~\subref{fig:ISI_W3} show $\isi/\s$ measured in 
simulation (circles, red online), approximated using 
homogeneity closure of 
Eq.~(\ref{eq:closure_ISI_old}) (light curve, green 
online), and approximated using the result of asymptotic 
analysis in Eq.~(\ref{eq:closure_ISI_new}) (dark curve, 
black online). The closures are evaluated using 
node and link quantities measured in the 
simulations. Light curves (green online) in 
figures~\subref{fig:ISI_WD1}, \subref{fig:ISI_WD2}, 
and~\subref{fig:ISI_WD3} show the relative error, 
Eq.~(\ref{eq:delta}), of the homogeneity approximation, 
while dark curves (black online) show the relative error 
due to the newly derived approximation. 
Cusps in the relative error curves correspond to the 
$\isi/\s$ homogeneity closure curve crossing through the 
curve measured in simulations. Simulations are performed 
on a network with $10^5$ nodes and $5\times10^5$ links 
following algorithm in~\cite{Gillespie}.} 
\label{fig:ISI_W}
 \end{figure}

 \begin{figure}[tb!]
 \subfigure[$w/r=10^{-1}$]{\includegraphics[]{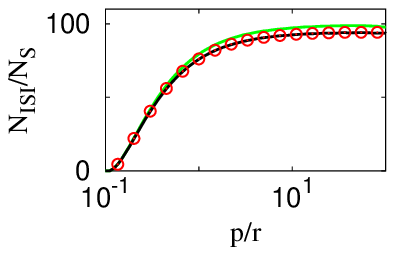}\label{fig:ISI_P1}}
 \subfigure[$w/r=10^{-1}$]{\includegraphics[]{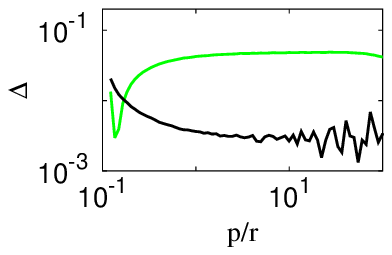}\label{fig:ISI_PD1}}
 \subfigure[$w/r=10^{0}$]{\includegraphics[]{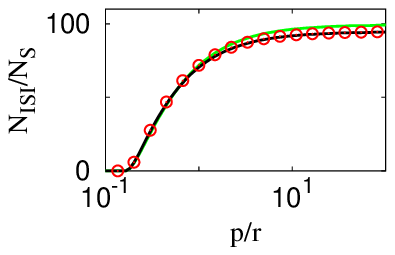}\label{fig:ISI_P2}}
 \subfigure[$w/r=10^{0}$]{\includegraphics[]{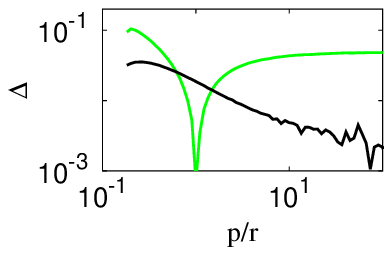}\label{fig:ISI_PD2}}
 \subfigure[$w/r=10^{1}$]{\includegraphics[]{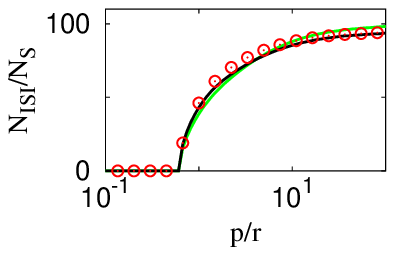}\label{fig:ISI_P3}}
 \subfigure[$w/r=10^{1}$]{\includegraphics[]{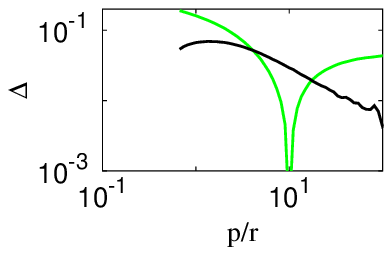}\label{fig:ISI_PD3}}
 \caption{Number of ISI triples per S-node as a function of infection rate, for
several rewiring rates: simulations compared to two approximations. The curves
and circles are defined as in Fig.~\ref{fig:ISI_W}, with the same network size
and number of links.}\label{fig:ISI_P}
 \end{figure}

Thus, we have derived a new closure of $\isi$: 
 \begin{align}
 &\frac{\isi}{\s}=
\frac{\is}{\s}+\frac{\is}{\s}\frac{\ssi}{\ss},\label{eq:closure_ISI_new}
 \end{align}
which relies on our ability to close the $\ssi$ term, and this brings 
us one step closer to finding an accurate closure of the 
mean-field description of the adaptive epidemic 
model~(\ref{eq:SIS_MF}). Curiously, the 
homogeneity closure of $\ssi$ in 
Eq.~(\ref{eq:closure_SSI_old}), together with 
Eq.~(\ref{eq:closure_ISI_new}), leads to the homogeneity 
closure in Eq.~(\ref{eq:closure_ISI_old}).  
Thus, as is suggested by Figs.~\ref{fig:ISI_W} 
and~\ref{fig:ISI_P}, where the measured values of 
$\ssi$ are used, improving the closure of $\ssi$ 
beyond the homogeneity assumption leads to 
improvement of the $\isi$ closure. In fact, 
Figs~\ref{fig:ISI_WD1},~\ref{fig:ISI_WD2},~\ref{fig:ISI_WD3} 
as well 
as~\ref{fig:ISI_PD1},~\ref{fig:ISI_PD2},~\ref{fig:ISI_PD3} 
show the relative deviation of the closure relations from 
the approximated quantity and suggest that the new 
approximation in Eq.~(\ref{eq:closure_ISI_new}) is 
superior to the relation in 
Eq.~(\ref{eq:closure_ISI_old}). Note that the only time 
the homogeneity closure appears to perform better is when 
it intersects the measured value of $\isi/\s$, and, 
therefore, its superiority over the performance of the 
new closure is rather coincidental. Further consideration 
of the results in Fig.~\ref{fig:ISI_W} shows that, as we 
move away from the derivation regime of slow 
rewiring rates, 
the performance of the new closure diminishes, though it 
is still superior to the old approximation. Predictably, 
as shown in Fig.~\ref{fig:ISI_P}, the performance of the 
new approximation improves for the larger values 
of infection rate, and outperforming 
the original closure even near the epidemic threshold.

\begin{figure}[tb!]
 \subfigure[]{\includegraphics[]{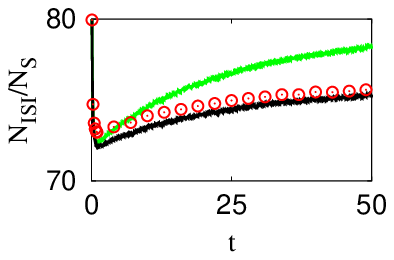}\label{fig:sis_time}}
 \subfigure[]{\includegraphics[]{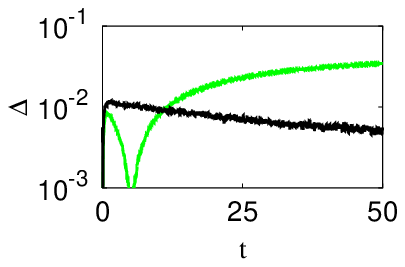}\label{fig:sis_time_triple}}
 \caption{Time evolution of ISI closure. 
Fig.~\subref{fig:sis_time} shows the time evolution of 
the number of ISI triples per S-node (circles, red 
online) and the approximate value as obtained from the 
relation in Eq.~(\ref{eq:closure_ISI_old}) (light curve, 
green online) and from Eq.~(\ref{eq:closure_ISI_new}) 
(dark curve, black online). 
Fig.~\subref{fig:sis_time_triple} shows the relative 
error due to the two approximations. The simulations are 
performed for $w=10^{-1}$, $p=10^0$, $r=1$, $10^5$ nodes 
and $5\times 10^5$ links. The initial network is a 
realization of an Erd\"{o}s-R\'{e}nyi random network; the 
state of each node is randomly assigned, with 90\% 
I-nodes and 10\% S-nodes. We take the average over 100 
dynamical realizations.
}
\label{fig:SIS_transient}
 \end{figure}

Finally, we test our new closure outside of the steady-state. Thus,
Fig.~\ref{fig:sis_time} compares the performance of the newly derived
approximation to that of the homogeneity approximation. We can see that, unlike
the homogeneity closure, the new closure follows the measured values of $\isi/\s$
very accurately. Furthermore, as shown in Fig.~\ref{fig:sis_time_triple}, the
closure of Eq.~(\ref{eq:closure_ISI_new}) performs better as the solution approaches the
steady-state.

 \section{Discussion}

 We presented an approach for closing a mean-field description of dynamical network systems. In our approach we
proposed exploiting the possible simplification of the heterogeneous mean-field description of the system in some
asymptotic regime. We applied this approach to two examples of adaptive networks: recruitment to a cause model
and a model of epidemic spread on an adaptive network. Using the two examples, we successfully developed
closures that perform as well as or better than the usual closures, which are based on the assumptions of
homogeneous distribution of nodes throughout the network.

 The closure we developed for the recruitment 
model showed significant improvement of the 
mean-field description over the one where all of 
the high order terms were approximated using the 
homogeneity closure. Not only do we see an 
improvement in the predicted levels of the 
recruited population; we also see greater 
consistency between the moment closure 
approximation and direct measurements of the 
closed terms. Thus, out of the three 
node-triple terms that we approximated, one 
showed significant improvement over the 
homogeneity based closure, and the sum of the 
remaining two triples proved to be consistent 
with the homogeneity closure.

In case of the epidemic model, the closure developed with the asymptotic approach also showed improvement over
the ad hoc, homogeneity based closure. The result of utilizing our approach was an improved moment closure
approximation for one of the terms, contingent on improvements of a closure for the other term, as confirmed
by the numerical simulations of the adaptive system.

It is important to note that the closures that we derived 
in some asymptotic regimes proved to be more accurate 
than the homogeneity closures even outside of the 
derivation limit. For example, even though in both cases 
the closures were derived at steady-state, they showed 
excellent results outside of the asymptotic parameter 
regime where the derivation took place, as well as during 
the transient state of the dynamical process. The 
additional benefit of using this approach is that it 
allows us to expect good performance of the closure at 
least in the limit where the derivation took place, more 
than can be said about any ad hoc moment closure 
approximation. However, the more rigorous
statements about the accuracy of this approach as well as 
the applicability of this approach to a more general 
class of network problems are left to the future 
investigations.

 \begin{acknowledgments}
 This work was supported by the Army Research 
Office, Air Force Office of Scientific Research, 
by Award Number R01GM090204 from the National 
Institute Of General Medical Sciences. MSS was also 
supported by 
the US National Science Foundation through grant 
DMR-1244666. The content is solely the 
responsibility of the authors and does not 
necessarily represent the official views of the 
National Institute of General Medical Sciences 
or the National Institutes of Health.

\end{acknowledgments}

\appendix
\section{Homogeneity Closure}\label{app:closure}

The homogeneity closure is based on the assumption that 
the probability of finding an R-node at the end of a link 
that stems from an S-node is independent of what else is 
in the neighborhood of that S-node and is given by 
$q=\rs/(\rs+\ss+\sn)$. 
In other words, the probability for the S-node to 
have $r$ of R-nodes in its neighborhood is assumed to 
be independent of the number of X-nodes in the 
neighborhood where X represents the other node 
types:
 \begin{align}
 P_{S;n_R|n_X}(r|x)=P_{S;n_R}(r)
 \end{align}
where
$P_{S;n_R|n_X}(r|x)$ is the probability
$r$ of the R-nodes are in the neighborhood of the
S-node
conditioned on the presence of $x$ of the X-nodes in
the neighborhood of that same S-node, and $P_{S;n_R}(r)$
is the probability distribution of the number of R-nodes in the neighborhood of S-nodes.
Note that this consideration is for $X\ne R$.

When the expected number of XSR triples per S-node is 
evaluated, the homogeneity assumption translates into the 
following relation:
 \begin{align}
 \begin{split}
 &\frac{N_{\text{XSR}}}{N_S} = \sum_{x,r} xr P_{S;n_R|n_X}(r|x)P_{S;n_X}(x)=\\
 &=\left[ \sum_{r} r P_{S;n_R}(r)\right]\left[\sum_{x} x P_{S;n_X}(x)\right]=\\
 &=\frac{N_{\text{XS}}}{\s}\frac{\rs}{\s}
 \end{split}
 \end{align}

In order to close the term describing the expected number of RSR triples per S-node, additional information
about the total degree distribution of S-nodes, $P_{S;n_D}$, is required.
We make an ad hoc assumption that the distribution is Poisson:
 \begin{align}
 P_{S;n_D}(d)=\frac{e^{-\mm} \mm^d}{d!},
 \end{align}
 which would be the case had the links between the nodes 
been formed in a random fashion. Here the mean of the 
distribution is known and given by $\mm=(\sn+\ss+\rs)/\s$.
The homogeneity assumption on the 
distribution of R-nodes implies that the probability that 
an S-node with total degree $d$ has $r$ of the R-nodes in 
its neighborhood, $P_{S;n_R|n_D}(r|d)$, is given by the 
binomial distribution:
 \begin{align}
 P_{S;n_R|n_D}(r|d)= {d \choose r} q^{r} (1-q)^{d-r}.
 \end{align}
The expected number of RSR node triples per S-node is now evaluated as follows:
 \begin{align}
 \begin{split}
 &\frac{N_{\text{RSR}}}{N_S} = \sum r^2 P_{S;n_R}(r)=\\
 &=\sum_{r,d} r^2 P_{S;n_R|n_D}(r|d) P_{S;n_D}(d)=\\
 &=\sum_{d} [(dq)^2+dq(1-q)] P_{S;n_D}(d)=\\
 &=q^2 (\mm^2+\mm)+q(1-q)\mm=\\
 &=q^2\mm^2 + q\mm=\\
 &=\left(\frac{\rs}{\s}\right)^2+\frac{\rs}{\s}.
 \end{split}
 \end{align}

The same approach leads to the homogeneity closure in the 
SIS model, where we replace recruiting nodes by infected 
nodes.

\section{Mean-Field Equations for Recruitment with Adaptation}\label{app:NSR}

The following mean-field equations are found by multiplying Eq.~(\ref{eq:master_RSR}) by
$k_1^{i_1}k_2^{i_2}k_3^{i_3}$ with
$i_1+i_2+i_3=0,1$:
\begin{subequations}
\begin{align}
&\dt  \n = \mu - \lA \n+\lB \s- \d \n, \label{eq:node_N}\\
&\dt  \s = \lA \n -\lB \s- \g\rs - \d \s, \label{eq:node_S}\\
&\dt  \r = \g \rs - \d \r, \label{eq:node_R}\\
\begin{split}
&\dt \nn = 2{\sigma \mu}\frac{\n}{\n+\s+\r}+2\lB \sn\\
& - 2(\lA+\d) \nn \label{eq:link_NN},
\end{split}\\
\begin{split}
&\dt \sn ={\sigma \mu}\frac{\s}{\n+\s+\r}+\lB \ss \\
&-\g \nsr-(\lA +\lB+2\d)\sn+ \lA \nn,
\end{split}\\
\begin{split}
& \dt \ss =-2\g \ssr+2\lA \sn\\
&-2(\lB+\d) \ss,
\end{split}\\
\begin{split}
&\dt \rn = \sigma \mu\frac{\r}{\n+\s+\r}+\g \nsr\\
&-(\lA +2\d+w)\rn +\lB \rs,
\end{split}\\
\begin{split}
& \dt \rs =- \g \rsr+ \g \ssr\\
&-(\lB+2\d) \rs +(\lA+w) \rn,
\end{split}\label{eq:link_RS}\\
& \dt \rr = 2\g \rsr  -2\d \rr.\label{eq:link_RR}
\end{align}\label{eq:NSR_MF}
\end{subequations}
At steady state, in the limit where $\g,\d,\mu/(\n+\s+\r) 
\ll w, \lA, \lB$, Eq.~(\ref{eq:node_S}) and 
Eq.~(\ref{eq:link_RS}) lead to the following relations:
 \begin{subequations}
 \begin{align}
 &\lA \n = \lB \s,\label{eq:nsr_limit1}\\
 &(\lA+w)\rn=\lB\rs.\label{eq:nsr_limit2}
 \end{align}
 \end{subequations}
 Note that here we do not take into consideration the 
fact that all of the terms $N_{X_1 X_2\ldots X_n}$ are 
functions of the system parameters. For example, we 
implicitly assume that $\lA, \lB$ 
and $w$ can be chosen large enough such that $\g \rs \ll 
\lB \s, \lA\n$ in the considered limit.

\section{Derivation of the $\nsr+\ssr$ closure.}\label{app:NSRSSR}

To obtain a closure of the $\nsr$ and $\ssr$ terms in the recruiting model, we consider the expression
 \begin{align}
 \begin{split}
 &\sum_{\k}\left(\frac{\rs}{\s}\frac{\dt\Ms}{\n}+\frac{\rn}{\n}
\frac{\dt\Mn}{\n}\right)\left(k_1 k_3 + k_2 k_3\right), \label{eq:HMFsum}
 \end{split}
 \end{align}
 which should be $0$ at steady state.  Discarding quantities proportional to 
parameters $\g,\d,\mu/(\n+\s+\r)$ (which are assumed small relative to other 
parameters), we find that the quantity in the first term of~(\ref{eq:HMFsum})
simplifies to
\begin{subequations}
\begin{align}
\begin{split}
&\sum_{\k}\left(\dt\Ms k_1 k_3 \right) = -\lB \nsr+\lA \nnr+\\
&+ \sum_{\k}\left[ \lA (k_1+1)k_1 k_3 \rhor[1] \right.
- \lA k_1^2 k_3 \Ms+\\
&+ \lB k_1(k_2+1)k_3 \rhor[2]-\lB k_1 k_2 k_3 \Ms+\\
&\left.+w \frac{\rn}{\s}k_1 k_3\rhor[9]-w \frac{\rn}{\s}k_1k_3\Ms\right]\\
&= -\lB \nsr+\lA \nnr - \lA \nsr + \lB \ssr + \\
&+w \frac{\rn}{\s} \sn
\end{split}
\end{align}
\end{subequations}
 using the fact that $\sum_{\k} k_1k_3 \Ms=\nsr$, $\sum_{\k} k_1k_3 \Mn=\nnr$, 
etc.  Other terms in~(\ref{eq:HMFsum}) sum similarly.

 We use Eq.~(\ref{eq:nsr_limit1})-(\ref{eq:nsr_limit2}) to eliminate 
parameters $\lambda_2,w$ from the expression~(\ref{eq:HMFsum}), replacing them 
with combinations of $\lambda_1$ and node and link variables. This yields
\begin{subequations}
\begin{align}
\begin{split}
&\frac{1}{\lambda_1 \n}\sum_{\k}\left[\dt\Ms (k_1 k_3 +k_2k_3)\right]=-\frac{\nsr}{\s} + \frac{\nnr}{\n}+\\
&+ (\frac{\rs}{\s}-\frac{\rn}{\n})\frac{\sn}{\s}-\frac{\ssr}{\s} + \frac{\snr}{\n}+\\
& +(\frac{\rs}{\s}-\frac{\rn}{\n})\frac{2\ss}{\s}\\
\end{split}
\end{align}
\end{subequations}
and
\begin{subequations}
\begin{align}
\begin{split}
&\frac{1}{\lambda_1 \n}\sum_{\k}\left[\dt\Mn (k_1 k_3 +k_2k_3)\right]=\frac{\nsr}{\s} -\\ &-\frac{\rs}{\s}\frac{\n}{\rn}\frac{\nnr}{\n}+\frac{\ssr}{\s}-\frac{\rs}{\s}\frac{\n}{\rn}\frac{\snr}{\n}
\end{split}
\end{align}
\end{subequations}

Combining these quantities as in~(\ref{eq:HMFsum}) and setting to $0$ for steady state finally yields
 \begin{align}
 \begin{split}
 &\left(\frac{\nsr}{\s}+\frac{\ssr}{\s}-\frac{\sn}{\s}\frac{\rs}{\s}-\frac{\ss}{\s}\frac{\rs}{\s}\right)\times\\
 &\times\left(\frac{\rn}{\n}-\frac{\rs}{\s}\right)=0,
 \end{split}
 \end{align}
 or
 \begin{align}
 \begin{split}
 &\frac{\nsr}{\s}+\frac{\ssr}{\s}=\frac{\sn}{\s}\frac{\rs}{\s}+\frac{\ss}{\s}\frac{\rs}{\s}.
 \end{split}
 \end{align}

\section{Mean-Field Equations for SIS with Adaptation}\label{app:SIS}
The mean-field equations generated from the heterogeneous mean-field equations~(\ref{eq:masterS})
and~(\ref{eq:masterI}) are the following:
 \begin{subequations}\label{eq:SIS_MF}
 \begin{align}
& \dt  \s = r\i - p \is, \label{eq:node_S2}\\
& \dt  \i = -r\i + p \is, \label{eq:node_I}\\
& \dt  \ss = -2p\ssi +2(r + w) \is, \label{eq:node_SS2}\\
\begin{split}
& \dt  \is = r\ii -p\isi + p\ssi\\
&-(r+w)\is,
\end{split}\label{eq:node_SI}\\
& \dt  \ii = -2r\ii +2p\isi\label{eq:node_II}
 \end{align}
 \end{subequations}
 The conservation
equations for nodes and links are:
 \begin{subequations}
 \begin{align}
 &\s+\i=N,\\
 &\ss+2\is+\ii=N \sigma,
 \end{align}
 \end{subequations}
 where $N$ is the total number of nodes, and $\sigma$ is the mean degree.
At steady-state the mean-field equations give the following relations used
in the body of the paper:
 \begin{subequations}
 \begin{align}
 &p\ssi=(r+w)\is, \label{eq:ss1}\\
 &p\isi=r\ii.\label{eq:ss2}
 \end{align}
 \end{subequations}

\bibliographystyle{apsrev}
\bibliography{Shkarbib}

\end{document}